\documentclass[twocolumn,10pt]{elsart4}
\usepackage{amssymb}
\usepackage{graphicx}
\usepackage{times}

\begin{document}

\pagestyle{empty}
\begin{frontmatter}

\vspace{-1cm}
\begin{center}
\small \slshape Conference on Turbulence and Interactions TI2006,
May 29 - June 2, 2006, Porquerolles, France
\end{center}

\title{Turbulence characteristics of the B\"odewadt layer
in a large shrouded rotor-stator system}
\author{S.\ Poncet$^{\dag,*}$, A.\ Randriamampianina$^{\dag}$}
\address{ \textnormal{$^{\dag}$IRPHE UMR 6594 CNRS - Universit\'es d'Aix-Marseille I et
II, Technop\^ole Ch\^ateau-Gombert, 49 rue F. Joliot-Curie BP 146,
13 384 Marseille c\'edex 13, FRANCE\\ $^{*}$Email:
poncet@irphe.univ-mrs.fr}}

\begin{abstract}
A three-dimensional direct numerical simulation (3D DNS) is
performed to describe the turbulent flow in an enclosed rotor-stator
cavity characterized by a large aspect ratio $G=(b-a)/h=18.32$ and a
small radius ratio $a/b=0.15$ ($a$ and $b$ the inner and outer radii
of the rotating disk and $h$ the interdisk spacing). Recent
comparisons with velocity measurements \cite{poncras} have shown
that, for the rotational Reynolds number $Re=\Omega b^2/\nu=95000$
($\Omega$ the rate of rotation of the rotating disk and $\nu$ the
kinematic viscosity of water) under consideration, the stator
boundary layer is 3D turbulent and the rotor one is still laminar.
Budgets for the turbulence kinetic energy are here presented and
highlight some characteristic features of the effect of rotation on
turbulence. A quadrant analysis of conditionally averaged velocities
is also performed to identify the contributions of different events
(ejections and sweeps) on the Reynolds shear stress.
\end{abstract}

\end{frontmatter}

\section*{Introduction}
Besides its primary concern to many industrial applications, the
rotor-stator problem has proved a fruitful means of studying
turbulence in confined rotating flows. This specific configuration
is indeed among the simplest ones where rotation brings significant
modifications to the turbulent field. The present paper is devoted
to the study of the turbulent flow in an enclosed high-speed
rotor-stator system of large aspect ratio. The flow is of Batchelor
type and belongs to the regime denoted IV by Daily and Nece
\cite{dail60}. These authors provided an estimated value for the
local Reynolds number at which turbulence originates with separated
boundary layers, $Re_r=\Omega r^2 / \nu=1.5 \times 10^5$ ($r$ the
radial location) for $G \leq 25$. However, experiments have revealed
that transition to turbulence can appear at a lower value of the
Reynolds number within the stator boundary layer (the B\"odewadt
layer), even though the flow remains laminar in the rotor boundary
layer (the Ekman layer). Recently, the 3D computed solution
presented here has been compared to velocity measurements performed
at IRPHE \cite{poncras}. It has been shown that, for the rotational
Reynolds number $Re=95000$, the B\"odewadt layer is turbulent and
the rotor one is still laminar. The purpose of this work is to
provide detailed data of the turbulent boundary layer along the
stator side in a large enclosed system ($G=18.32$).

\section*{Numerical approach}
The numerical approach is based on a pseudo-spectral technique using
Chebyshev polynomials in the radial and axial directions with
Fourier series in the azimuthal direction associated with a
semi-implicit second order time scheme. An efficient projection
method is introduced to solve the pressure-velocity coupling. A grid
resolution composed of $N \times M \times K = 300 \times 80 \times
100$ respectively in radial, axial and azimuthal directions has been
used, with a dimensionless time step $\delta t = 2.75 \times
10^{-3}$. Numerical computations have been carried out on NEC SX-5
(IDRIS, Orsay, France).

\section*{Results and discussion}
\begin{figure}[ht]
\begin{center}
\includegraphics[width=0.3\textwidth]{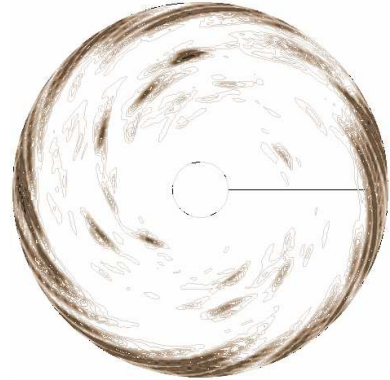}
\end{center}
\caption{Iso-contours of the instantaneous turbulence kinetic energy
within the stator boundary layer.} \label{fg1}
\end{figure}

The detailed description of the mean field and of the axial
variations of the Reynolds stress tensor is given in \cite{poncras}.
Nevertheless, we recall the main results. A good agreement has been
obtained between the 3D solution and the experimental data, whereas
the axisymetric solution leads to a steady laminar flow. The flow is
of Batchelor type: the two boundary layers developed on each disk
are separated by a central rotating core characterized by a quasi
zero radial velocity and a constant tangential velocity. The
B\"odewadt layer along the stator is centripetal, three-dimensional
as the Townsend structural parameter is lower than the limit value
$0.15$, and turbulent with turbulence intensities increasing from
the axis to the periphery of the cavity. On the contrary, the Ekman
layer on the rotor is centrifugal and laminar. The turbulence is
observed by the formation of turbulent spots developing along spiral
arms towards the periphery of the cavity, as seen in figure
\ref{fg1} from the instantaneous iso-values of the turbulence
kinetic energy within the stator boundary layer.

\subsection*{Turbulence kinetic energy budgets}
The balance equation for the turbulence kinetic energy $k$ can be
written:
\begin{eqnarray} \label{TKEB}
A = P + D^T + D^{\nu} + \Pi - \epsilon
\end{eqnarray}

\noindent with the advection term $A$, the production term $P$, the
turbulent diffusion $D^T$, the viscous diffusion $D^{\nu}$, the
velocity-pressure-gradient correlation $\Pi$ and the dissipation
term $\epsilon$.

\vspace{0.4cm}
\begin{figure}[ht]
\begin{center}
\includegraphics[width=0.35\textwidth]{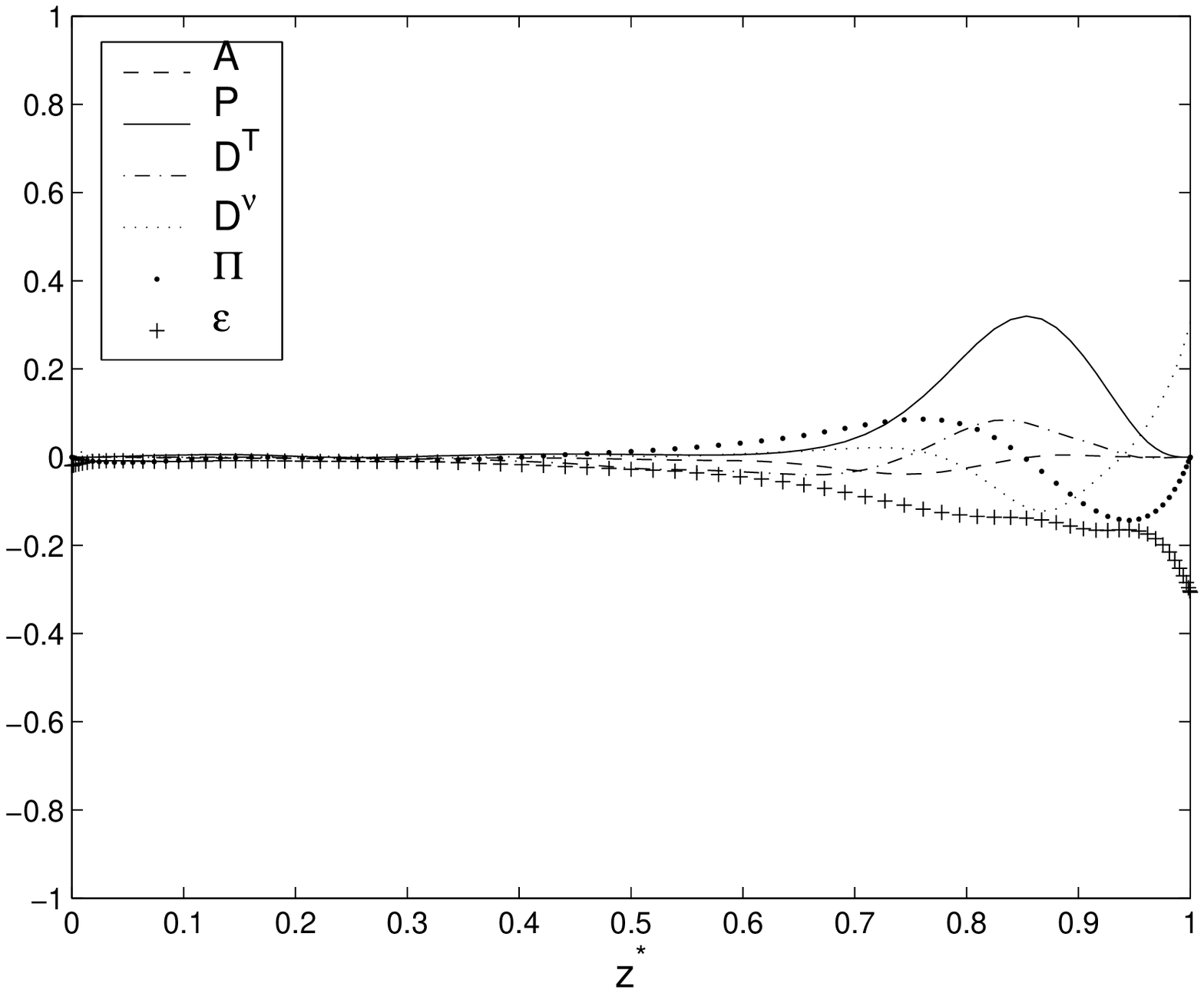}
\\
(a)
\\
\includegraphics[width=0.35\textwidth]{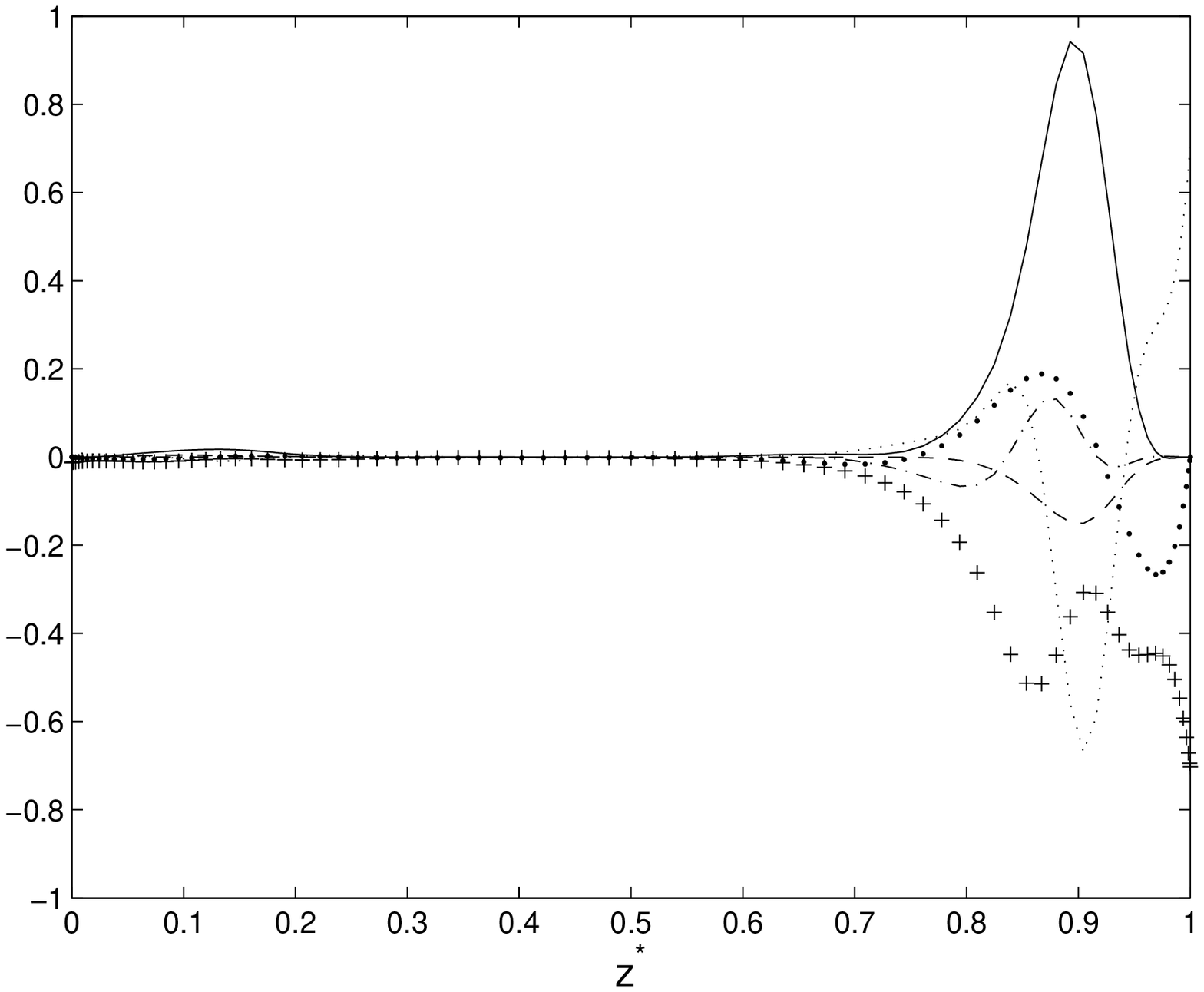}
\\
(b)
\end{center}
\caption{Budgets for the turbulence kinetic energy $k/(\Omega b)^2$
at: (a) $r/b=0.56$, (b) $r/b=0.8$.} \label{bilankz}
\end{figure}

Figures \ref{bilankz}a and \ref{bilankz}b show the axial variations
along $z^*=z/h$ of the different terms involved in the transport
equation (\ref{TKEB}) at two radial locations. It is clearly seen
that all these terms vanish towards the rotor side ($z^*=0$),
confirming the laminar nature of this zone up to the stator boundary
layer ($z^*=1$). At the stator wall, the viscous diffusion balances
the dissipation as well known in 3D turbulent boundary layer. Within
the B\"odewadt layer, even though some interaction between the
different terms involved is observed, the major contributions come
from the production, the dissipation and the viscous diffusion
terms. The production is balanced by the dissipation and the viscous
diffusion, the level of which increases in association with the
thickening of the boundary layer towards the periphery. The
production increases with increasing radius as already observed with
the levels of the normal Reynolds stresses \cite{poncras}. The
maximum of the production term is obtained at the wall coordinate
$z^+=z v_{\tau}/\nu=12$ ($v_{\tau}=((\nu \partial
V_{\theta}/\partial z)^{2}+(\nu \partial V_{r}/\partial
z)^{2})^{1/4}$ the total friction velocity and $z$ the axial
coordinate) for $r/b=0.56$ and at $z^+=12.5$ for $r/b=0.8$, which
confirms the approximately self-similar behavior of the B\"odewadt
layer. The levels of the viscous diffusion increase when moving
towards the outer casing, where the highest turbulence intensities
prevail. It indicates that viscous effects still play an important
role in the turbulence towards these regions, which does not allow
for a distinct delineation of the viscous sublayer.  This indicates
also the weak nature of the turbulence obtained at this Reynolds
number.

\subsection*{Conditional-averaged quadrant analysis}
\begin{figure}[ht]
\begin{center}

\vspace{1.6cm}

\includegraphics[width=0.35\textwidth]{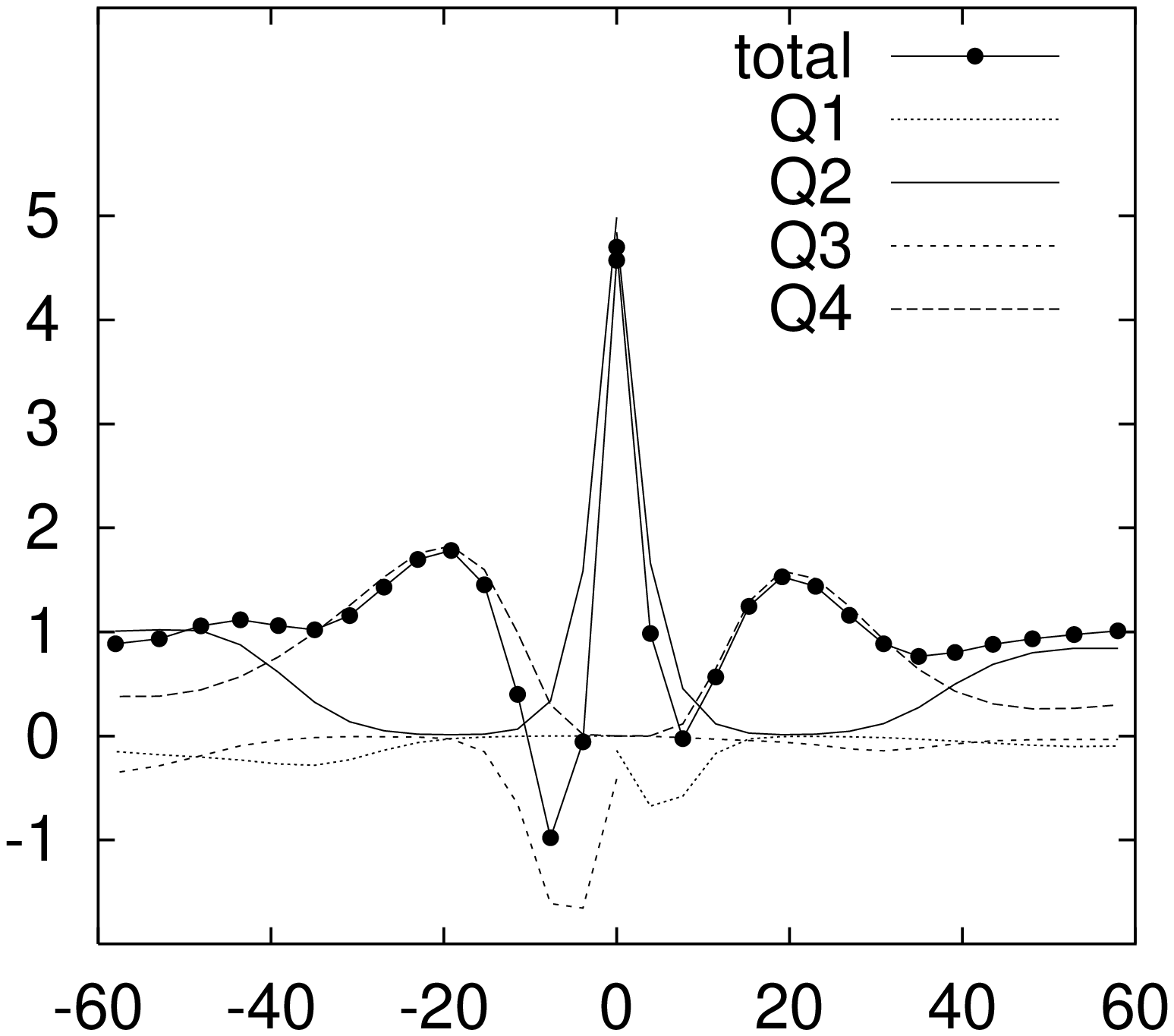} \\
(a)

\vspace{1.6cm}

\includegraphics[width=0.35\textwidth]{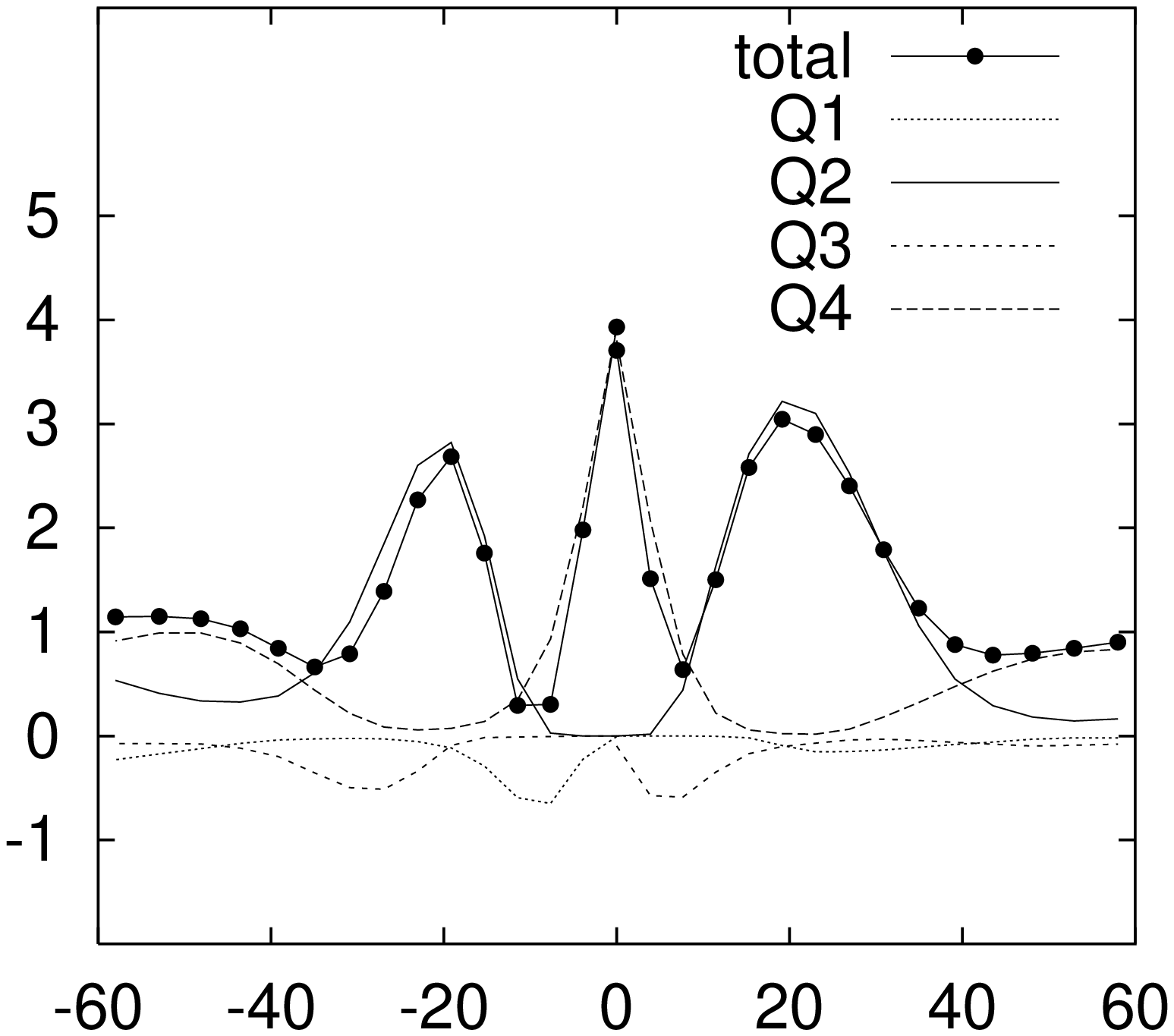}\\
(b)
\end{center}
\caption{Variation with $-\Delta r^+$ of the conditionally averaged
Reynolds shear stress at $z^+=17$ in the vicinity of (a) a strong
ejection $<v'_{\theta}v'_z \mid strong$
$ejection>/<v'_{\theta}v'_z>$ and (b) a strong sweep
$<v'_{\theta}v'_z \mid strong$ $sweep>/<v'_{\theta}v'_z>$.}
\label{vwsweep}
\end{figure}

To gain a better insight on the near-wall structure of the turbulent
boundary layer along the stator side, a conditional-averaged
quadrant analysis is performed to identify the contributions of
intense intermittent bursting events (ejections and sweeps) on the
Reynolds shear stress producing vortical structures. It corresponds
to four subdivisions of the fluctuation field according to the
combination of the tangential $v'_{\theta}$ and axial $v'_z$
velocity fluctuations \cite{litte94,Kang98}. Following the
definitions given in \cite{lygr01} in a fixed frame, a strong sweep
is associated with $-v'_{\theta}v'_z>\beta v'_{\theta,rms}
v'_{z,rms}$ and $v'_z<0$ (quadrant Q4) and a strong ejection with
$-v'_{\theta}v'_z> \beta v'_{\theta,rms} v'_{z,rms}$ and $v'_z>0$
(quadrant Q2). In the first quadrant Q1, $v'_{\theta}>0$ and
$v'_z>0$, while in the third quadrant Q3, $v'_{\theta}<0$ and
$v'_z<0$. The quadrant analysis is applied at $z^+=17$ corresponding
to the location of the maximum value of the turbulent shear stress.
We have also considered the value $\beta=2$ to determine strong
events, as used in \cite{litte94,Kang98}. We display in figures
\ref{vwsweep}a and \ref{vwsweep}b the variations with $\Delta r^+= r
\pm r^+$ ($r^+$ the wall coordinate in the radial direction) of the
conditionally averaged Reynolds shear stress normalized by the
unconditionally ensemble averaged Reynolds shear stress
$<v'_{\theta}v'_z>$ near a strong ejection (fig.\ref{vwsweep}a) and
a strong sweep (fig.\ref{vwsweep}b). The contributions of each
quadrant are also presented. The figures \ref{vwsweep}a and
\ref{vwsweep}b clearly show that the ejection (Q2) and sweep (Q4)
quadrants contribute much more to the Reynolds shear stress
production than the two other quadrants. On the other hand, it seems
that the weakness of the turbulence in the present simulation
accentuates the features observed in previous works. The results
obtained support the conclusions of Littell and Eaton \cite{litte94}
and Lygren and Andersson \cite{lygr01}, in contrast with the
findings of Kang et $al.$ \cite{Kang98}: the asymmetries observed
are dominated by Reynolds stress-producing coherent structures
(sweep and ejection). Lygren and Andersson \cite{lygr01} concluded
that clockwise vortices contribute much more to the Reynolds shear
stress than counter-clockwise vortices. The same behavior applies in
the presence of a sweep event. In this case, the levels of the
surrounding ejections approach the strong sweep level and are even
slightly beyond the fixed criterion condition $\beta=2$, as seen in
figure \ref{vwsweep}b, while the levels of sweeps around a strong
ejection are less important (fig.\ref{vwsweep}a), in agreement with
the results of \cite{lygr01}. Case 1 vortices, having induced
near-wall velocity in the direction of the crossflow, are found to
be the major source of generation of special strong events.

\section*{Conclusion}
DNS calculations have been used to describe the turbulent boundary
layer along the stator side in a large enclosed rotor-stator cavity.
For the rotational Reynolds number under consideration $Re=9.5
\times 10^4$, the B\"odewadt layer is 3D turbulent, whereas the
Ekman layer on the rotor is still laminar. The transition to
turbulence is associated with the onset of localized turbulent spots
spiraling up along the stator side. The turbulence kinetic energy
budgets have revealed that production is the major contribution with
a maximum obtained at $z^+ \simeq 12$ independently of the radial
location, confirming the self-similar behavior of the B\"odewadt
layer. The results of the quadrant analysis support the conclusions
proposed by \cite{litte94,lygr01}. The asymmetries observed by these
authors have been clearly detected and the analysis of conditionally
averaged streamwise and wall-normal velocity components confirms
that these asymmetries mainly arise from the contributions of
quadrants Q2 and Q4, responsible for the generation of ejection and
sweep events. Therefore, Case 1 vortices are found to be the major
source of generation of special strong events.


\begin{thebibliography}{9}
\bibitem{poncras} S. Poncet \& A. Randriamampianina "\'Ecoulement turbulent dans une cavit\'e rotor-stator ferm\'ee de grand rapport d'aspect,"
C.R. M\'ecanique, vol. 333, pp. 783-788, 2005.

\bibitem{dail60} J.W. Daily \& R.E. Nece "Chamber dimension effects on
induced flow and frictional resistance of enclosed rotating disks,"
ASME J. Basic Eng., vol. 82, pp. 217-232, 1960.

\bibitem{litte94} H.S. Littell \& J.K. Eaton "Turbulence characteristics of the boundary layer on a rotating
disk," J. Fluid. Mech., vol. 266, pp. 175-207, 1994.

\bibitem{Kang98} H.S. Kang, H. Choi \& J.Y. Yoo "On the modification of the near-wall coherent structure in a
three-dimensional turbulent boundary layer on a free rotating disk,"
Phys. Fluids, vol. 10 (9), pp. 2315-2322, 1998.

\bibitem{lygr01} M. Lygren \& H.I. Andersson "Turbulent flow between a rotating and a stationary disk,"
J. Fluid Mech., vol. 426, pp. 297-326, 2001.

\end{thebibliography}
\end{document}